\documentclass[11pt]{article}

\usepackage[T1]{fontenc}
\usepackage[utf8]{inputenc}
\usepackage{lmodern}
\usepackage{amsmath, amsthm, amssymb,float}
\usepackage{amsfonts}
\usepackage{graphicx}
\usepackage{dcolumn}
\usepackage{bm}
\usepackage{textcomp}
\usepackage[normalem]{ulem}
\usepackage[toc,page]{appendix}

\usepackage{color}
\usepackage{graphicx}
\usepackage{dcolumn}
\usepackage{bm}
\usepackage{amssymb}
\usepackage{amsmath}
\usepackage{mathrsfs}
\usepackage{color}

\usepackage{epsfig}
\usepackage{textcomp}
\usepackage{epstopdf}
\title{Photon Acceleration in Magnetized Plasma:\\
A Mechanism for Fast Radio Bursts}

\author{
S.~V.~Bulanov$^{1}$,
G.~Grittani$^{1}$,
M.~Lama\v{c}$^{1}$,
P.~Valenta$^{1}$,
S.~S.~Bulanov$^{2}$,\\
T.~Zh.~Esirkepov$^{3}$,
G.~Gregori$^{4}$,
B.~Russell$^{5}$,
A.~Thomas$^{6}$,
A.~Vanthieghem$^{7}$
}

\date{\today}

\begin{document}

\maketitle

\begin{center}
\small
$^{1}$ELI Beamlines Facility, Extreme Light Infrastructure ERIC,\\
Za Radnici 835, 25241 Dolní Břežany, Czech Republic

\vspace{0.4em}

$^{2}$Lawrence Berkeley National Laboratory, Berkeley, California 94720, USA

\vspace{0.4em}

$^{3}$National Institutes for Quantum Science and Technology (QST),\\
Kansai Institute for Photon Science, Kizugawa, Kyoto 619-0215, Japan

\vspace{0.4em}

$^{4}$University of Oxford, Oxford, United Kingdom

\vspace{0.4em}

$^{5}$Department of Astrophysical Sciences, Princeton University,\\
Princeton, New Jersey 08544, USA

\vspace{0.4em}

$^{6}$Gerard Mourou Center for Ultrafast Optical Science,\\
University of Michigan, Ann Arbor, Michigan 48109, USA

\vspace{0.4em}

$^{7}$Sorbonne Université, Observatoire de Paris, Université PSL,\\
CNRS, LERMA, Paris, France
\end{center}

\begin{abstract}
We propose a mechanism for fast radio bursts (FRBs) based on photon
acceleration by relativistic shocks in highly magnetized electron--positron
plasmas, as expected in magnetar magnetospheres. Density modulations at the
shock front create relativistically moving refractive-index perturbations that
transform low-frequency electromagnetic precursors into enhanced high-frequency
radiation. The observed frequencies, durations, and energetics of FRBs are
shown to be consistent with magnetic-field strengths, shock Lorentz factors,
and spatial scales expected in magnetar magnetospheres.
\end{abstract}


\section{Introduction}

Fast radio bursts (FRBs) are bright, millisecond-duration radio transients of predominantly extragalactic origin, first reported by Lorimer et al. \cite{Lorimer} (see also review article by Lorimer et al. \cite{LorimerRev}), who discovered a highly dispersed burst consistent with propagation through the intergalactic medium. Since then, numerous FRBs have been detected, exhibiting a wide range of temporal structures, spectral features, and polarization properties, as well as both repeating and apparently non-repeating sources (see Petroff et al. \cite{Petroff2022}. Despite rapid observational progress, the physical mechanism responsible for their intense, coherent radio emission remains not fully understood and continues to pose a number of open questions (see Lyubarsky \cite{Lyubarsky2014}; Beloborodov \cite{Beloborodov2017}; Popov \cite{Popov2018}; Zhang \cite{Zhang2020}; Lyutikov \cite{Lyutikov2021}; Zhang \cite{Zhang2023}).

The extreme brightness temperatures inferred for FRBs require coherent emission mechanisms operating in highly dynamic plasma environments. A broad class of models invokes strongly magnetized compact objects, particularly magnetars, whose large magnetic energy reservoirs and episodic energy release can naturally account for the observed energetics. In these models, magnetic reconnection or crustal activity launches relativistic outflows into the surrounding plasma. Alternatively, coherent emission may be produced at relativistic shocks, such as those driven by magnetar flares or in gamma ray burst (GRB)-like environments, where radiation is generated in either the upstream or downstream region of the shock.

Several mechanisms for coherent emission in such systems have been considered. Early work by Colgate and Noerdlinger \cite{Colgate1971} examined coherent radiation from expanding plasma shells. More recent studies have focused on relativistic shock physics, including the generation and amplification of precursor electromagnetic waves through plasma instabilities and ion–electron coupling (e.g., Iwamoto et al. \cite{Iwamoto2}). The “moving mirror” model of Yalinewich and Pen \cite{Yalinewich} describes the frequency upshifting and amplification of radiation reflected from relativistically moving plasma structures as a result of a double Doppler effect (Einstein \cite{Einstein}). In addition, Deng and Wu \cite{Deng} invoked the interaction of electromagnetic pulses with refractive-index modulations propagating at relativistic velocities to explain the circular polarization observed in some fast radio bursts. A comprehensive review of FRB phenomenology and theoretical models is provided by Zhang \cite{Zhang2023}.

In this paper, we present a physical model of FRBs in which a relativistic parallel shock wave in an electron–positron pair plasma produces refractive index modulations propagating at relativistic velocity. At the same time, the shock generates beams of high-energy particles which, via beam–plasma instability, excite electromagnetic precursors in the upstream region that propagate with a group velocity lower than that of the shock.
Reflection of these relatively low-frequency precursors from the shock front leads to the generation of an intense, coherent electromagnetic burst, with the frequency upshifted according to the theory of relativistic mirror (Einstein \cite{Einstein}). The theoretical framework we employ is based on the photon acceleration mechanism, which describes the interaction of electromagnetic waves with relativistically moving refractive index modulations within the geometric optics approximation (Wilks et al. \cite{Wilks}). 

Here, we extend this mechanism to the case of magnetized electron-positron pair plasma. In many of the scenarios of coherent electromagnetic radiation generation in space, relativistic plasma waves and strong electromagnetic fields play a central role in mediating energy transfer to radiation. However, the influence of strong background magnetic fields on photon acceleration processes in such environments has not been fully explored. Magnetization can significantly alter the dispersion properties of plasma waves and their nonlinear interaction with electromagnetic radiation.

In this work, we investigate photon acceleration in magnetized plasma and demonstrate that strong magnetic fields can enhance both the frequency upshift and efficiency of this process. We show that relativistic plasma disturbances, such as those associated with shock waves, can convert low-frequency, low-amplitude perturbations into intense, high-frequency electromagnetic pulses. In magnetar magnetospheres, where the magnetic field is extremely large and plasma flows can become relativistic during flares, the conditions for efficient photon acceleration are naturally satisfied. Low-frequency electromagnetic perturbations arising from magnetospheric activity or upstream precursor waves can be swept up by relativistic shocks and converted into short, intense, high-frequency pulses. The resulting emission is coherent, highly amplified, and temporally compressed, consistent with the observed properties of repeating FRBs.

\section{Photon accelerator}

Within the framework of photon acceleration concept (see Wilks et al. \cite{Wilks}), the propagation of a short-wavelength electromagnetic wave packet in plasma is described using the geometric optics (WKB) approximation. In this approach, the wave packet is treated as a quasi-particle characterized by its central position  $\mathbf{x}(t)$ and wave vector  $\mathbf{k}(t)$. 
The ray equations for the wave packet within 1D approximation can be written in the form
   \begin{equation}
    \label{eq1}
      \dot{x} =\partial_k \omega  \quad {\text{and}} \quad \dot{k} =-\partial_k \omega ,
   \end{equation}	
where the wave frequency  $\omega(k,x,t)$ is determined by the local dispersion equation for the waves propagating in the plasma. Here and below a dot stands for the time derivative. These equations have a Hamiltonian form with the Hamiltonian equal to the wave frequency, $\mathcal{H}=\omega(k,x,t)$. If the plasma parameters depend on $x$  and  $t$ only through the combination
   \begin{equation}
    \label{eq2}
      \xi =x-\beta_M c t,
   \end{equation}	
i.e., the wave frequency profile (it is given by $\omega(k,\xi)$) propagates rigidly with velocity $v_M=\beta_M c$. In this case, the system admits an integral of motion, usually called the Jacobi integral, which is defined by expression
   \begin{equation}
    \label{eq3}
      \mathcal{H}-k c \beta_M =J.
   \end{equation}	
Thus, the quantity
   \begin{equation}
    \label{eq4}
      \omega-k c \beta_M =J
   \end{equation}	
is conserved along the ray trajectory. The Jacobi integral represents the conservation of the wave frequency $\omega^{\prime}$  in the frame of reference moving with the plasma disturbance,  $J=\omega^{\prime}/\gamma_M$, where  $\gamma_M=1/\sqrt{1-\beta_M^2}$. This invariant forms the basis for describing photon acceleration in nonstationary plasmas, as it directly relates changes in wave frequency to variations in the plasma properties and the velocity of the underlying disturbance.

\section{Electromagnetic waves in magnetized electron-positron plasma}

To specify the dependence of the wave frequency on the plasma parameters we consider the case of the electron-positron plasma, which is relevant to magnetar magnetospheres where FRB are generated. For both the right and left circularly polarized electromagnetic waves (R- and L-modes) propagating in electron-positron plasma along the constant magnetic field  $B$ the dispersion equation is
   \begin{equation}
    \label{eq5}
    \frac{k^2 c^2}{\omega^2} = 1-\frac{\omega_p^2}{\omega^2-\omega_B^2}, 
   \end{equation}	
where the Langmuir and Larmor frequencies equal $\omega_p=\sqrt{4 \pi n_{\pm}e^2/m_e}$ 
and $\omega_B=eB/m c$ for  $n_{\pm}=2n_e$ (e.g. see Zhang \cite{{Zhang2023}} and references therein). 
Here $e$ and $m_e$ are the electron charge and mass, respectively,
and $c$ is the speed of light in vacuum.

In the case when the linearly polarized electromagnetic wave propagates across the constant magnetic field, there are two modes. For the ordinary (O-mode) wave with the electric field parallel to the magnetic field the dispersion equation has the same form as for the wave in unmagnetized plasma 
   \begin{equation}
    \label{eq6}
    \frac{k^2 c^2}{\omega^2} = 1-\frac{\omega_p^2}{\omega^2},
   \end{equation}	
and for the extraordinary (X-mode) wave the dispersion equation is given by Eq. (\ref{eq5}).

As noted above, in this paper we analyze the photon accelerator properties in magnetized electron-positron pair plasma as inspired by the demand to develop a physical mechanism for fast radio bursts. The case of the photon accelerator in magnetized electron-ion plasma has been considered by Mendonca \cite{Mendonca2001}, Eliezer et al. \cite{Eliezer2005}, and Bulanov et al. \cite{Bulanov2026}. The effect of strong magnetic fields on high power gamma ray flash generation in the laser pulse interaction with solid targets is studied by Hadjisolomou et al. \cite{Hadjisolomou2023}, where it is shown that the efficiency of the laser energy conversion to the gamma radiation energy is substantially increased when the laser-magnetized target configuration corresponds to the X-wave case. 
The electromagnetic wave can propagate in plasma provided the square of the refraction index is positive. According to Eqs. (\ref{eq5}) and (\ref{eq6})  the transparency regions are beyond the frequency intervals 
$\omega_B<\omega<\sqrt{\omega_p^2+\omega_B^2}$  and $\omega<\omega_p$, respectively. In the high frequency limit when the wave frequency is well above the Larmor frequency, $\omega_B$, the wave behaves as in the case 
of unmagnetized plasma (or as the O-mode). In the opposite low frequency limit when $\omega << \omega_B$  the dispersion equation 
   \begin{equation}
    \label{eq7}
    \frac{k^2 c^2}{\omega^2} = 1+\frac{\omega_p^2}{\omega_B^2} 
   \end{equation}	
describes the Alfven-like mode with 
   \begin{equation}
    \label{eq8}
        \omega=k c \frac{\omega_B}{\omega_{UH}} 
        = k \frac{v_a}{\sqrt{1+v_a^2/c^2}}=k c \sqrt{\frac{\sigma_e}{1+\sigma_e^2}}
   \end{equation}	
Here 
   \begin{equation}
   \omega_{UH}=\sqrt{\omega_p^2+\omega_B^2}
      \end{equation}	
   is the “upper hybrid” frequency, 
   \begin{equation}   
   v_a=B/\sqrt{4\pi n_{\pm}m_e}
      \end{equation}	
   is the Alfven velocity, calculated for the density of electron-positron pair plasma and the electron mass, and
   \begin{equation}
   \sigma_e=B^2/4\pi n_{\pm}m_e c^2
     \end{equation}	
   is so-called electron magnetization parameter. In an electron–positron plasma, the low-frequency electromagnetic mode is Alfvén-like and non-dispersive, the whistler mode does not exist, there is no difference between the R-mode (circular right polarized wave) and the L-mode (circular left polarized wave) propagating along the magnetic field, due to exact mass symmetry. The group velocity of the Alfven-like wave equals $v_g=\partial_k \omega=c \omega_B/\omega_{UH}$.

\section{Phase plane for photon acceleration by O-wave or/and in unmagnetized plasma}

We consider the phase space of the photon accelerator in an unmagnetized plasma by O-wave or/and in unmagnetized electron-positron pair plasma, when in the Jacobi integral given by Eq. (\ref{eq4}) the relationship between the wave frequency and wavenumber is determined by the dispersion equation (\ref{eq6}). Contours of constant Jacobi integral define trajectories in the $(\xi,\omega)$  plane. For a given dependence of the plasma frequency $\omega_p(\xi)$  these contours are defined as follows:
   \begin{equation}
    \label{eq9}
        \omega-\beta_M\sqrt{\omega^2-\omega_p^2}=J.
   \end{equation}	

These phase trajectories describe the evolution of the wave packet as it propagates through the nonuniform and nonstationary plasma, providing a convenient representation of photon acceleration in terms of coupled variations of the wave frequency $\omega$  and coordinate $\xi$.

\begin{figure}
    \centering
\includegraphics[width=0.75\columnwidth]{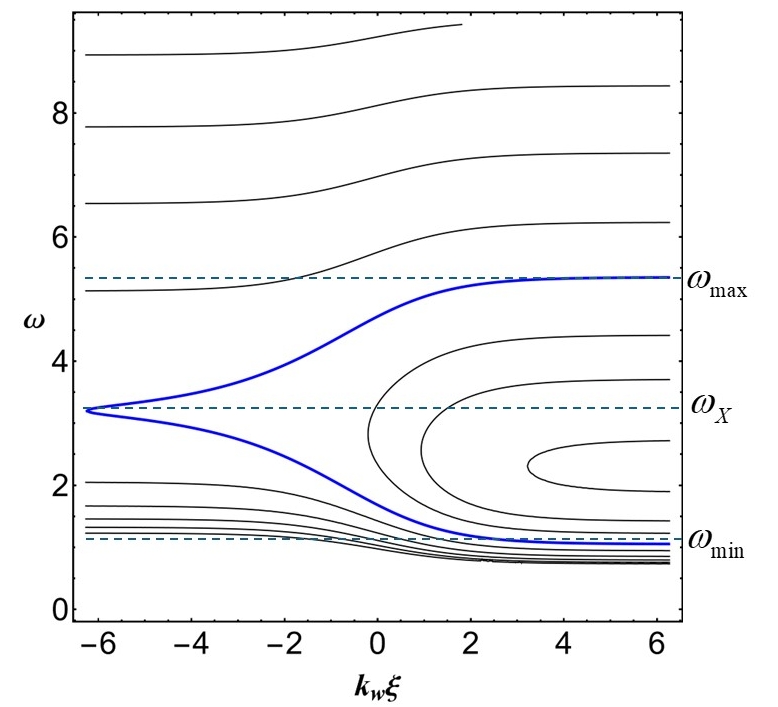}
    \caption{
    Contours of constant value of the Jacobi integral in the
    $(k_w \xi,\omega)$ plane for $\beta_M=0.95$, $k_w=1$,
    and $\omega_p(-\infty)=1$.
    Here $\omega_{\min}$ and $\omega_{\max}$ are the minimum and maximum
    frequencies on the separatrix (blue curves), given by
    Eqs.~(\ref{eq14}) and (\ref{eq15}), respectively.
    The critical point of the separatrix is
    $(-\infty,\omega_X)$.
    }
    \label{Fig1}
\end{figure}

To illustrate photon acceleration in the case of an electromagnetic wave interacting with refractive-index modulations in the form of a shock wave, we present in Fig. \ref{Fig1} contour lines corresponding to constant values of the Jacobi integral, for a spatial dependence of the plasma frequency squared (i. e., proportional to electron-positron density)
   \begin{equation}
    \label{eq10}
        \omega_p^2=\omega_p^2(-\infty)
        \left(\beta_M\sqrt{1-\delta+\delta\frac{\exp{-k_w}}{1+\exp{-k_w\xi}}}\right).
   \end{equation}	

According to this expression the Langmuir frequency monotonously depends on the coordinate $\xi$ being equal to 
$\omega_p(-\infty)$  at $\xi\to -\infty$  and to  $\omega_p(-\infty)\sqrt{1-\delta}$ for $\xi\to +\infty$. 
The shock wave front thickness is approximately equal to $1/k_w$. Here the parameter $\delta$ characterizes 
the change in plasma density at the shock wave.


As follows from the analysis of constant value Jacobi integral in the  $(k_w\xi,\omega)$ variables, the trajectories are partitioned into two classes: bounded (reflected) and unbounded (transient). The region of reflected trajectories is delimited by the separatrix (shown by the blue curve in Fig. \ref{Fig1}), which constitutes the level set of the Jacobi integral passing through the critical equilibrium point $(-\infty,\omega_X)$ . The branches of the separatrix intersect at this unstable fixed point, where the group velocity of the electromagnetic wave packet coincides with the velocity of the refractive index modulation,
\begin{equation}
    \label{eq11}
        \left.\partial_k\omega\right|_{\xi \to -\infty}=c\frac{\sqrt{\omega_X^2-\omega^2_p(-\infty)}}{\omega_X}=\beta_M c.
   \end{equation}	
   This condition defines the frequency corresponding to the critical point in phase space,
\begin{equation}
    \label{eq12}
        \omega_X=c\frac{\omega_p(-\infty)}{\sqrt{1-\beta_M^2}}.
   \end{equation}	
which is indicated in Fig.\ref{Fig1} by the horizontal dashed line. Using this expression we can find the Jacobi integral value on the separatrix. It is equal to
\begin{equation}
    \label{eq13}
       J_X= \omega_p(-\infty)\sqrt{1-\beta_M^2}.
   \end{equation}	
This invariant determines the frequency shift of the reflected electromagnetic wave. For an incident pulse whose parameters lie on the lower branch of the separatrix, the reflected trajectory maps onto the upper branch, where the frequency attains its maximal value. The minimal and maximal frequencies along the separatrix are given by
\begin{equation}
    \label{eq14}
       \omega_{\max}=\frac{\omega_p(-\infty)+\beta_M\sqrt{\omega_p(-\infty)^2-\omega_p(+\infty)^2}}{\sqrt{1-\beta_M^2}} 
   \end{equation}	
   and
 \begin{equation}
    \label{eq15}
       \omega_{\min}=\frac{\omega_p(-\infty)-\beta_M\sqrt{\omega_p(-\infty)^2-\omega_p(+\infty)^2}}{\sqrt{1-\beta_M^2}},
   \end{equation}  
respectively. Accordingly, the ratio of the maximal to minimal frequency,
 \begin{equation}
    \label{eq16}
      \frac{\omega_{\max}}{\omega_{\min}}=\frac{\omega_p(-\infty)+\beta_M\sqrt{\omega_p(-\infty)^2-\omega_p(+\infty)^2}}
      {\omega_p(-\infty)-\beta_M\sqrt{\omega_p(-\infty)^2-\omega_p(+\infty)^2}},
   \end{equation} 
for $\omega_p(-\infty)/\omega_p(+\infty)\gg 1$  is approximately equal to
 \begin{equation}
    \label{eq17}
      \frac{\omega_{\max}}{\omega_{\min}}\approx\frac{1+\beta_M}{1-\beta_M}\approx 4 \gamma_M^2
   \end{equation} 
in agreement with Einstein’s formula. Here, $\gamma_M=1/\sqrt{1-\beta_M^2}$  denotes the Lorentz factor corresponding to the velocity  $\beta_M c$ of the refractive index modulation.

Finally, the characteristic photon acceleration length (dephasing length), determined by the phase mismatch accumulated along the trajectory, is given by
 \begin{equation}
    \label{eq18}
     l_{acc} =\frac{2\pi}{k_w(1-\beta_M)}.
   \end{equation} 

   \section{Photon acceleration in magnetized electron-positron plasma}

   As noted above, in a magnetized electron–positron plasma the dispersion equation (\ref{eq6}), which describes the dependence of the wavenumber on frequency, exhibits distinct transparency regions depending on the wave polarization and propagation direction. For a linearly polarized electromagnetic wave propagating perpendicular to the magnetic field, with the electric field perpendicular to the magnetic field (the X-mode), as well as for circularly polarized waves propagating parallel to the magnetic field, the transparency region lies outside the frequency interval  $\omega_B<\omega<\omega_{UH}$. 
   
In these cases, the Jacobi integral can be written in the form
 \begin{equation}
    \label{eq19}
    \omega-\beta_M \omega\sqrt{1-\frac{\omega_p^2}{\omega^2-\omega_B^2}}=J.
   \end{equation} 

\begin{figure}
    \centering
   \includegraphics[width=0.75\columnwidth]{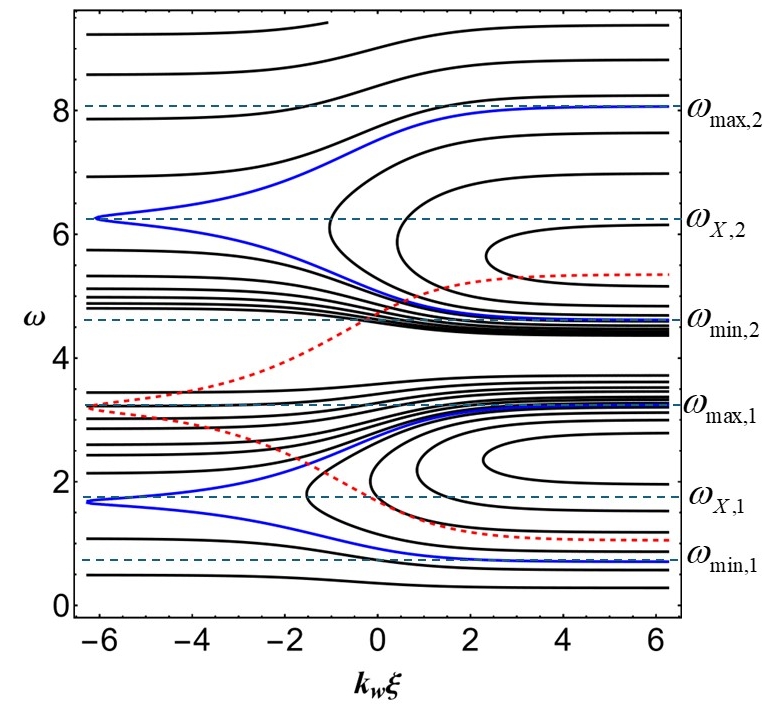}
    \caption{ 
   Contours of constant value of the Jacobi integral dependent on $\omega$  and $\xi$  
   for  $\beta_M=0.95$,  $k_w=1$, $\omega_p(-\infty)=1$, $\omega_B=4$ and  $\delta=0.5$. Here  $\omega_{\min,1}$,   $\omega_{\max,1}$,    $\omega_{X,1}$ and  $\omega_{\min,2}$,   $\omega_{\max,2}$,    $\omega_{X,2}$   are the given by Eqs. (\ref{eq24}, \ref{eq27}) and (\ref{eq29}) minimal, maximal frequency on the separatrices (blue curves) whose critical points are   $(\omega_p(-\infty),\omega_{X,1})$ and   $(\omega_p(-\infty),\omega_{X,2})$. The red dotted curves show the position of the separatrix for the same parameters, but in a non-magnetized electron-positron plasma (see Fig. {\ref{Fig1}}). }
    \label{Fig2}
\end{figure}

   The contours of constant Jacobi integral, corresponding to the spatial dependence of the plasma frequency given by Eq. (\ref{eq10}), are presented in Fig. \ref{Fig2}. The inclusion of the magnetic field leads to a qualitative restructuring of the underlying phase space. In particular, the frequency interval in which the plasma is opaque (i.e.  $\omega_B<\omega<\omega_{UH}$) acts as a forbidden band, dividing the phase plane into two disconnected regions. 
   
   Within each region, the phase portrait exhibits a critical equilibrium point (for the plasma density profile determined by Eq. (\ref{eq10}) they are located at  $\xi\to -\infty$), whose stable and unstable manifolds form separatrices (shown as blue curves). These separatrices define the boundaries between qualitatively different classes of wave packet propagation: bounded (reflected) trajectories and unbounded (transmitting) trajectories. As a result, each subdomain possesses its own separatrix-bounded region of reflection embedded within a domain of transit trajectories, giving rise to a nontrivial phase–space topology. For comparison, the red dotted curves show the separatrix for the same parameters, but in a non-magnetized electron-positron plasma (see Fig. \ref{Fig1}).

To analyze the topology of the phase space and determine the position of the separatrices and the maximum and minimum frequencies of reflected electromagnetic waves, it is necessary to calculate the dependence of the group velocity of the wave on either the wave number or the wave frequency. Solving the dispersion equation (\ref{eq5}) for the wave frequency, we obtain
\begin{equation}
    \label{eq20}
    \omega=\sqrt{\frac{k^2c^2+\omega_{UH}^2\pm\sqrt{k^4c^4+\omega_{UH}^2+2k^2c^2(\omega_p^2-\omega_B^2)}}{2}}.
   \end{equation} 
   
\begin{figure}
    \centering
   \includegraphics[width=0.75\columnwidth]{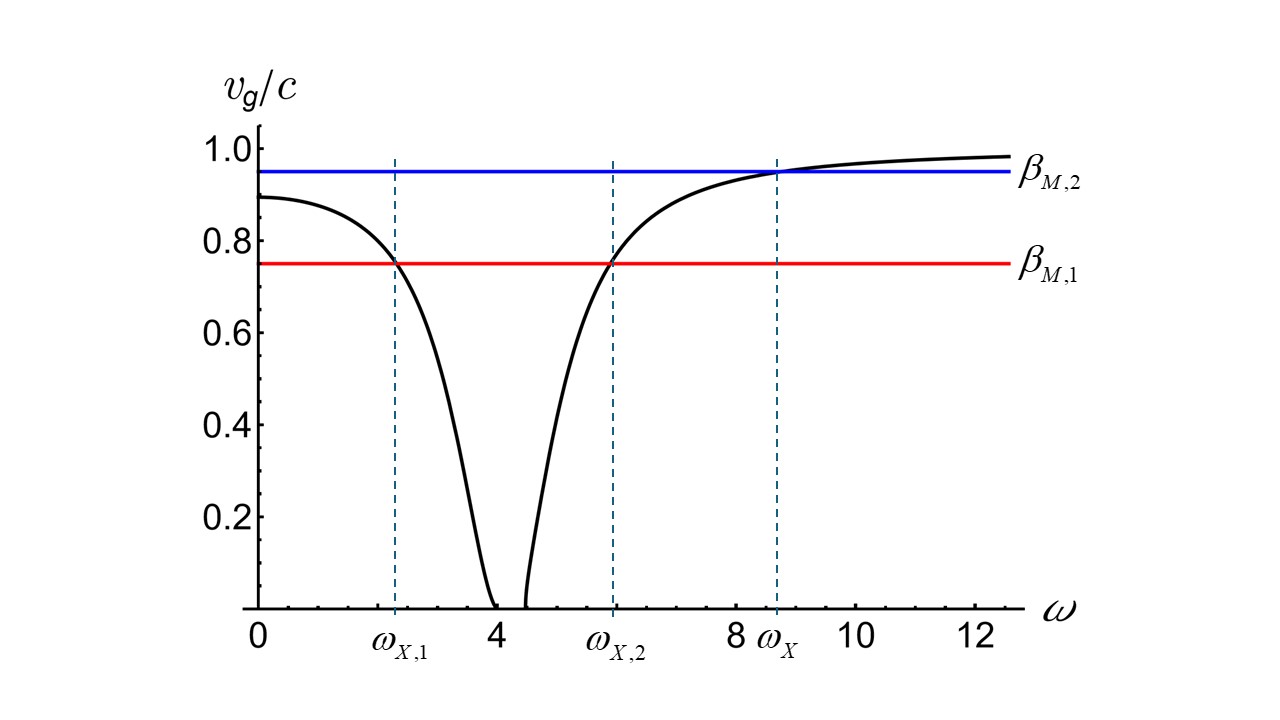}
    \caption{ 
    Group velocity versus wave frequency for $\omega_p=2$  and $\omega_B=4$  (black curves). The red and blue lines correspond to   $\beta_M=0.75$ and $\beta_M=0.95$, respectively. Here $\omega_{X,1}$,  $\omega_{X,2}$, and  $\omega_{X}$  denote the critical-point frequencies at which the group velocity equals propagation velocity of the refractive-index modulations. }
    \label{Fig3}
\end{figure}

Differentiating this function with respect to the wave number yields the group velocity $v_g=\partial_k\omega$ as a function of the wave number.
\begin{equation}
    \label{eq21}
    v_g(k)=kc^2\frac{\omega_B^2-\omega_p^2-k^2c^2\pm\sqrt{k^4c^4+\omega_{UH}^2+2k^2c^2(\omega_p^2-\omega_B^2)}}{\sqrt{2[k^2c^2+\omega_{UH}^2\pm\sqrt{k^4c^4+\omega_{UH}^2+2k^2c^2(\omega_p^2-\omega_B^2)}]}}.
   \end{equation} 
This dependence is multivalued. Therefore, it is more convenient to express the group velocity as a function of the wave frequency. It is given by
\begin{equation}
    \label{eq22}
    v_g(\omega)=
    \frac{1}
    {\partial_{\omega}k}=c\frac{(\omega^2-\omega_b^2)^{3/2}\sqrt{\omega^2-\omega_{UH}^2}}
    {(\omega^2-\omega_b^2)^2+\omega_P^2\omega_b^2}.
   \end{equation} 
   Its dependence on frequency is shown in Fig. \ref{Fig3} for $\omega_p=2$  and $\omega_B=4$  (black curves). An opacity gap appears in the interval $\omega_B<\omega<\omega_{UH}$, where the electromagnetic wave cannot propagate. The red and blue lines correspond to the normalized velocities of the refractive-index modulation, $\beta_M=0.75$ and $\beta_M=0.95$, respectively. Their intersections with the curves $v_g(\omega)$ determine the critical-point frequencies  $\omega_{X,1}$,  $\omega_{X,2}$, and  $\omega_{X}$.
   
Depending on the magnetic field strength and the values of the refractive-index modulation velocities, there may be either two critical points (at $\omega_{X,1}$ and  $\omega_{X,2}$   as in Fig. \ref{Fig2}) or a single one  $\omega_{X}$.

\subsection{Phase plane parameters in the lower subdomain}

To find the value of the lower critical frequency, we look for the asymptotic (at $\omega\to 0$ ) dependence of the group velocity given by equation (\ref{eq22}), which looks as follows.
\begin{equation}
    \label{eq23}
   \left. v_g(\omega)\right|_{\omega \to 0}=
  c\left[  \frac{\omega_B}{\omega_{UH}}-\frac{3\omega_p^2 \omega^2}{2\omega_B\omega_{UH}^3}
  \right].
   \end{equation} 

Using this expression and condition  $v_g(\omega_{X,1})=\beta_mc$  and assuming that $\beta_M<\omega_B/\omega_{UH}$  (we do not consider here the case when  $\beta_M>\omega_B/\omega_{UH}$, for which no separatrix exists in the lower subdomain of the phase plane), we find the lower critical frequency. It is approximately equal to 
\begin{equation}
    \label{eq24}
   \omega_{X,1}=
   \frac{\sqrt{2(\omega_p^2(-\infty)+\omega_B^2)(\omega_B^2-\beta_M\omega_B\sqrt{\omega_p^2(-\infty)+\omega_B^2})}}
   {\sqrt{3}\omega_p(-\infty)}
     \end{equation} 

In the limit $\beta_m\approx \omega_b/\omega_{UH} \to 1$ and $\omega_p/(\omega_B\sqrt{1-\beta_M}) \to 0$, from Eq. (\ref{eq22}) and condition $v_g(\omega_{X,1})=\beta_Mc$ the critical frequency is approximately equal to 
 \begin{equation}
    \label{eq25}
   \omega_{X,1}=\omega_B-\frac{2}{\sqrt{1-\beta_M}}.
     \end{equation} 
This allows us to calculate the value of Jacobi’s integral on the separatrix in the lower subdomain of the phase plane and find maximal frequency on the separatrices. As a result, we can see that the maximal frequency is about the Larmor frequency (this is also clearly seen in Fig. \ref{Fig2}) while the minimal frequency is approximately equal to 
\begin{equation} \label{eq26}
 \omega_B\frac{1-\beta_M}{1+\beta_M}.
\end{equation}

\subsection{Phase plane parameters in the upper subdomain}

To analyze the topology of the upper subdomain of phase plane using expression (\ref{eq22}) it is convenient to rewrite the equation for the critical frequency, for which the group velocity equals $v_g(\omega_{X,2})=\beta_M c$, in the form
 \begin{equation}
    \label{eq27}
  (1-\beta_M^2)X^4-\omega_p^2(-\infty)X^3-2\beta_M^2\omega_B^2\omega_p^2(-\infty)X^2-\beta_M^2\omega_B^4\omega_p^4(-\infty)=0,
     \end{equation} 
where 
\begin{equation} \label{eq26a}
X=\omega_{X,2}^2-\omega_B^2.
\end{equation} 
For $1-\beta_M^2 \to 0$ asymptotic solution of this equation is given by  
\begin{equation} \label{eq26b}
X\approx\frac{\omega_p^2(-\infty)}{1-\beta_M^2},
 \end{equation} 
i.e. the frequency at the critical point is equal to
 \begin{equation}
    \label{eq28}
  \omega_{X,2}\approx\sqrt{\omega_B^2+\frac{\omega_p^2(-\infty)}{1-\beta_M^2}},
      \end{equation} 
which gives for the value of Jacobi integral on the separatrix 
 \begin{equation}
    \label{eq29}
  J_{X,2}=(1-\beta_M^2)\sqrt{\omega_B^2+\frac{\omega_p^2(-\infty)}{1-\beta_M^2}}
      \end{equation}
and for the frequency upshift
 \begin{equation}
    \label{eq30}
\Delta\omega=2\sqrt{\omega_B^2+\frac{\omega_p^2(-\infty)}{1-\beta_M^2}}.
      \end{equation}
Comparing these expressions with Eqs. (\ref{eq14}–\ref{eq17}), we see that magnetic-field effects can enable the acceleration of photons to higher energies than in an unmagnetized plasma. 

\section{Beam instability in magnetized electron-positron plasma as a source of electromagnetic precursors}

We expect that activity in the magnetar magnetosphere can, in principle, generate low-frequency electromagnetic perturbations, for example through starquakes or other mechanisms, which may subsequently be amplified through their interaction with the shock wave. As an example of one such mechanism, we consider below electromagnetic precursors propagating into the upstream region.

Electromagnetic precursors in the upstream region generated by beams of particles accelerated by relativistic shock wave have been observed in a number of computer experiments starting from publication by Iwamoto et al. \cite{Iwamoto1} to more recent papers by Marcowith et al. \cite{Marcowith}, Plotnikov and Sironi \cite{Plotnikov2019}; Iwamoto et al. \cite{Iwamoto2}; Vanthieghem and Levinson \cite{Vanthieghem2025}; Mahlman et al. \cite{Mahlman2026}. Various instabilities of electrostatic electromagnetic types have been considered analytically and with computer simulations (Kazimura et al.  \cite{Kazimura1998}; Medvedev and Loeb \cite{Medvedev1999}; Beloborodov \cite{Beloborodov2013}; Golant, Vanthieghem, Grošelj, Sironi \cite{Golant25}). To describe the instability of electron-positron beam in magnetized plasma we invoke the theory developed by Achatz, Lesch and Schlikeiser \cite{Achatz1990} and Achatz and Schlikeiser \cite{Achatz1993}. According to this work, there are two types of unstable modes: electrostatic and electromagnetic. The development of the electrostatic mode leads to excitation of waves whose electric field is directed along the motion of the electron–positron beam. Such waves are not further intensified through interaction with a relativistic plasma mirror.

We expect that activity in the magnetar magnetosphere can, in principle, generate low-frequency electromagnetic perturbations that may subsequently be amplified through their interaction with the shock wave. As an example, we consider below the electromagnetic precursors emerging in the upstream region.

In contrast, the electromagnetic mode leads to the generation of low-frequency electromagnetic perturbations whose electric field is perpendicular to the direction of the electron–positron beam motion. These perturbations can be intensified and therefore may be considered as candidates for the origin of Fast Radio Bursts. The dispersion equation governing the unstable electromagnetic mode can be derived as follows. The dispersion equation in the frame of reference where the electron-positron pair beam is at rest is
\begin{equation}
    \label{eq31}
k^{\prime 2} c^2-\omega^{\prime 2}=-\frac{\omega_b^{\prime 2}\omega^{\prime 2}}{\omega^{\prime 2}-\omega_B^{2}},
      \end{equation}
where $\omega_b^{\prime}=\sqrt{4 \pi n^{\prime}_{\pm,b} e^2/mLe}$  is the Langmuir frequency with $n^{\prime}_{\pm,b}$  being the beam density in the boosted frame of reference. In the laboratory frame of reference, where the beam moves along the $x$ direction with velocity $v_b=\beta_b c$, we have for the frequency and wave number 
\begin{equation}
    \label{eq32}
\omega=\gamma_b(\omega^{\prime}+k^{\prime}c\beta_b), \quad \text{and} \quad k=\gamma_b(k^{\prime}+\omega^{\prime}\beta_b/c).
      \end{equation}
Here $\gamma_b=1/\sqrt{1-\beta_b^2}$ is the beam relativistic gamma-factor.

Including the electron–positron pair plasma contributions in the laboratory frame yields 
(Achatz, Lesch, and Schlickeiser, \cite{Achatz1990}; see also Lifshitz and Pitaevskii \cite{Lifshitz1981}, where the case of electron-ion plasma is considered):

\begin{equation}
    \label{eq33}
k^{2} c^2-\omega^{2}=-\frac{\omega_p^{2}\omega^{2}}{\omega^2-\omega_B^{2}}-\frac{\omega_b^{2}(\omega-kc \beta_b)^2}{\gamma_b((\omega-kc \beta_b)^2-\omega_B^{2}/\gamma_b^2)}.
      \end{equation}
We can rewrite this equation as
\begin{equation}
    \label{eq34}
\left(k^{2} c^2-\omega^{2}+\frac{\omega_p^{2}\omega^{2}}{\omega^{2}-\omega_B^{2}}\right)\left((\omega-kc \beta_b)^2-\frac{\omega_B^{2}}{\gamma_b^2}\right)=-\frac{\omega_b^2 (\omega-kc \beta_b)^2}{\gamma_b}.
      \end{equation}
The first factor on the left-hand side of the equation corresponds to the electromagnetic wave in the electron-positron pair plasma, and the second to the beam modes; the term on the right-hand side describes the interaction of these modes. By finding the intersection point of these branches, we can calculate the instability growth rate over a wide range of parameters. However, a full analysis of the instability behavior is beyond the scope of this article. 

In the context of the present work, we are interested in the low-frequency limit of unstable non-resonant mode excitation arising from the interaction of an electron–positron beam with an electron–positron plasma. This limit (i.e. $k\to 0$) corresponds to the approximation in which  $(\omega-k c\beta_b)^2<<\omega_B^2/\gamma_b^2$, and $\omega<<\omega_B$. Under these conditions, dispersion equation (\ref{eq34}) takes the following form:
\begin{equation}
    \label{eq35}
\left(k^{2} c^2-\omega^{2}\frac{\omega_{UH}^{2}}{\omega_B^2}\right)\omega_B^{2}=\omega_b^{2} \gamma_b (\omega-kc \beta_b)^2.
      \end{equation}
It describes the interaction of the Alfven-like wave whose dispersion equation is given by Eq. (\ref{eq8}) and very low frequency beam mode. Its solution yields 
\begin{equation}
    \label{eq36}
\omega=k c\frac{\beta_b\gamma_b\omega_{b}^{2}\pm\sqrt{\gamma_b\omega_b^2(\omega_B^2-\beta_b^2\omega_{UH}^2)+\omega_B^2\omega_{UH}^2}}{\gamma_b\omega_b^2+\omega_{UH}^2}.
      \end{equation}
From Eqs. (\ref{eq34}) and (\ref{eq36}) follows that the instability conditions are 
\begin{equation}
    \label{eq37}
k<\frac{\omega_B}{c\beta_b\gamma_b} \quad \text{and} \quad 
\omega_b>\frac{\omega_{UH}\omega_B}{\sqrt{\gamma_b(\beta_b^2\omega_{UH}^2-\omega_B^2)}}\approx \frac{\omega_{UH}\omega_B}{\sqrt{\gamma_b}\omega_p}.
      \end{equation}

According to Eq. (\ref{eq36}) the group velocity of excited beam-plasma mode (it is given by $\partial_k\Re{[\omega]}$)
\begin{equation}
    \label{eq37a}
v_g=c\beta_b\frac{\omega_{b}^{2}\gamma_b}{\omega_b^2\gamma_b+\omega_{UH}^2}
      \end{equation}
is smaller than the velocity of the electron-positron beam, which is equal to $c \beta_b$, and it can be smaller than the shock wave velocity $c \beta_M$.      

The growth rate, $\Gamma=\Im [\omega]$, is equal to
\begin{equation}
    \label{eq38}
\Gamma=k c\frac{\sqrt{\gamma_b\omega_b^2(\beta_b^2\omega_{UH}^2-\omega_B^2)-\omega_B^2\omega_{UH}^2}}{\omega_b^2\gamma_b+\omega_{UH}^2}
\approx kc \beta_b \frac{\omega_p}{\sqrt{\gamma_b}\omega_b}.
      \end{equation}
The non-resonant electromagnetic mode is unstable within the range of wave numbers
\begin{equation}
    \label{eq39}
0<k<k_{\max}=\frac{\omega_B}{c\beta_b\gamma_b}.
      \end{equation}

Using the relations derived above, we can estimate the maximum growth rate of the instability, which is approximately given by
\begin{equation}
 \label{eq40}
\Gamma_{\max}= \omega_b\frac{\omega_B}{\omega_{UH}\sqrt{\gamma_b}}     
\end{equation}

\begin{figure}
    \centering
    \includegraphics[width=\linewidth]{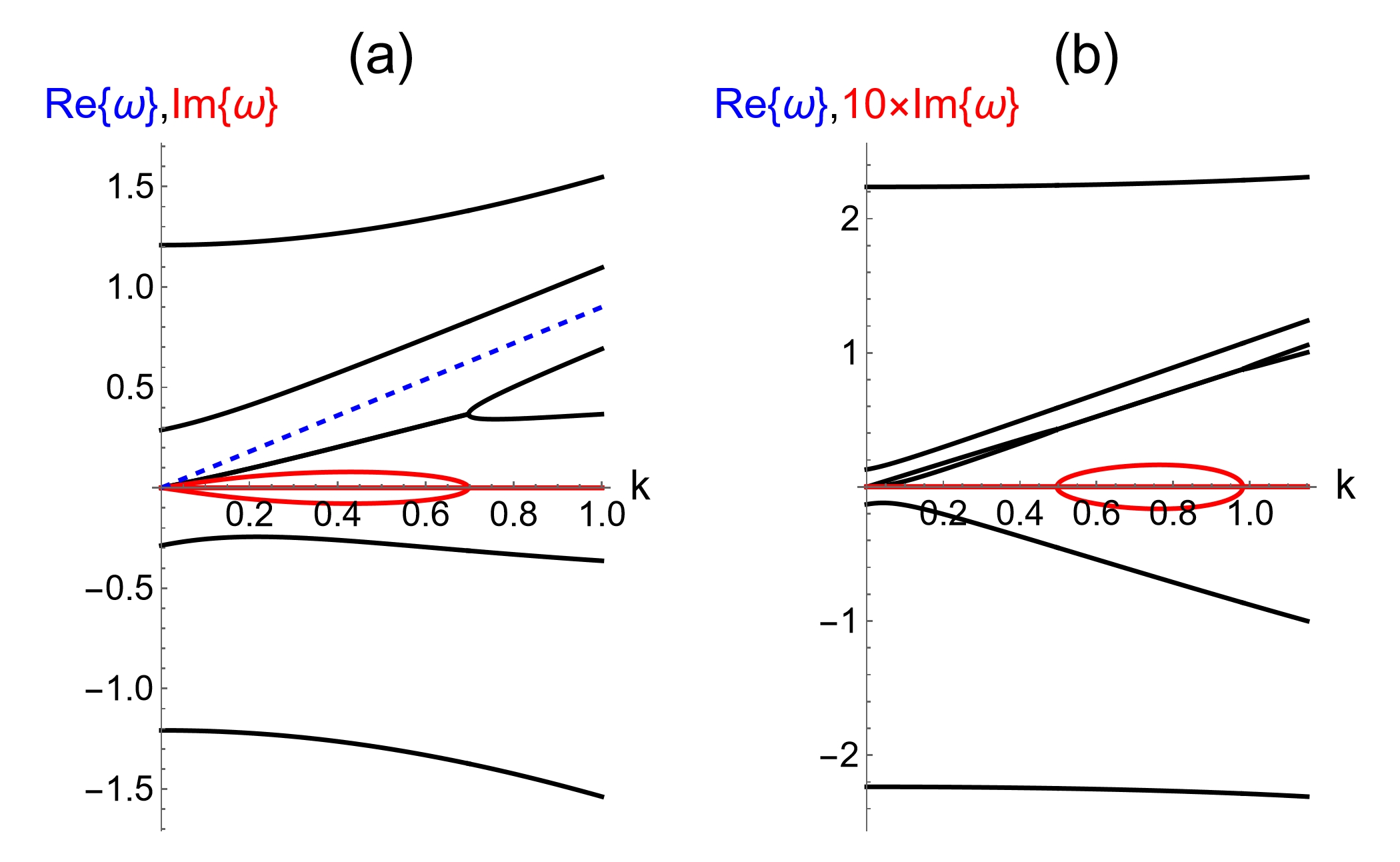}
    \caption{ 
    A pattern of the dependence of the real and imaginary parts of the frequency. The case of the non-resonant long-wavelength mode is shown in Frame a) for the normalized parameters $\omega_p=1$, $\omega_B=1$, $\omega_b=0.7$, and $\beta_b=0.98$. The case of the cyclotron unstable mode presented in Frame b) for the normalized parameters $\omega_p=1$, $\omega_B=2$, $\omega_b=0.75$, and $\beta_b=0.9$ with the growth rate, $\Gamma=\Im[\omega]$, multiplied by a factor of 10 for better visibility. The mode frequency and wave number are normalized to $\omega_b$   and  $\omega_b/c$, respectively. The real part of the frequency is shown in blue, while the imaginary part is shown in red.}
    \label{Fig4}
\end{figure}

A complete dependence of the real and imaginary parts of the unstable-mode frequency is presented in Fig. \ref{Fig4}. The results are obtained by numerically solving the dispersion equation (\ref{eq34}) for the normalized parameters corresponding to the low-frequency, non-resonant unstable mode: $\omega_p/\omega_B=1$, $\omega_b/\omega_B=0.7$, and $\beta_b=0.98$, corresponding to $\gamma_b=5$, as shown in frame a). In frame b), we present the real and imaginary parts of the frequency for the cyclotron instability. In this case, the growth rate, $\Gamma=\Im[\omega]$, is multiplied by a factor of 10 for better visibility.

The mode frequency and wave number are normalized to $\omega_b$   and  $\omega_b/c$, respectively. The real part of the frequency is shown in blue, while the imaginary part is shown in red.

\begin{figure}
    \centering
    \includegraphics[width=\linewidth]{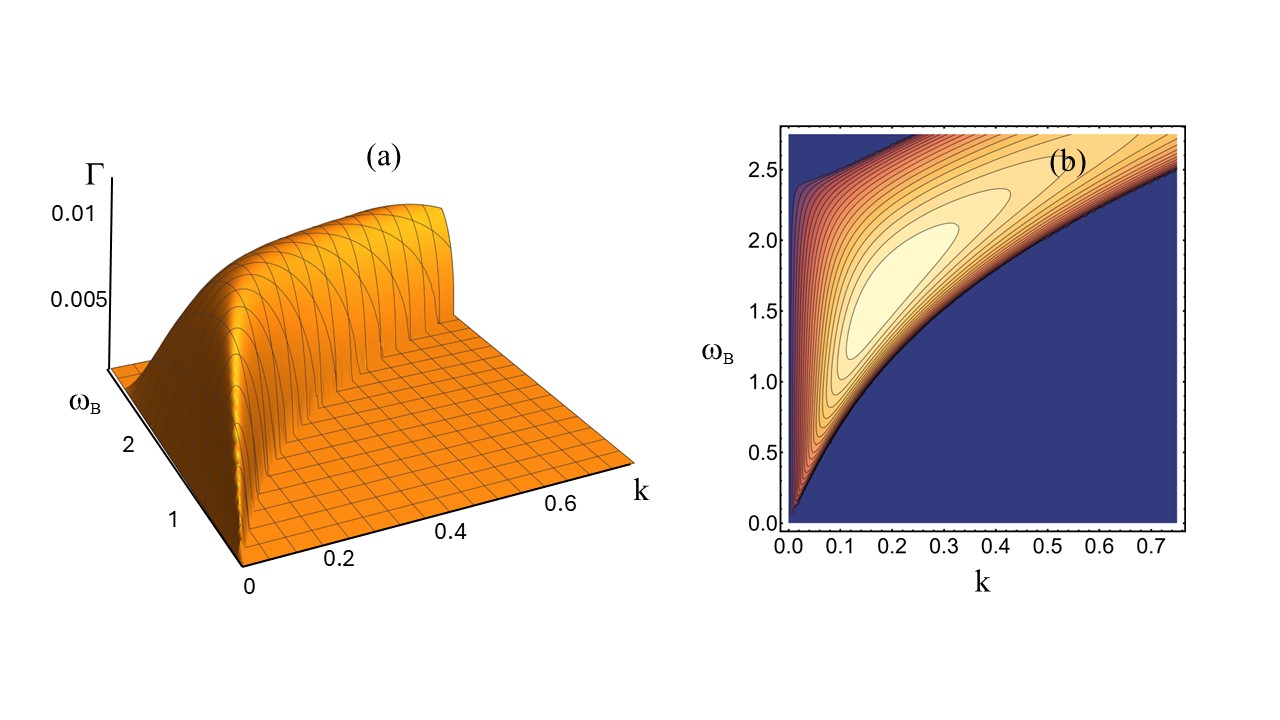}
    \caption{ 
    The growth rate on non-resonant (long wavelength,  $k<\omega_B/c\beta_b\gamma_b$) and short-wavelength modes  $k<\omega_B/c\beta_b\gamma_b$  for $\omega_p=1$, $\omega_b=0.7$  and $\beta_b=0.9999$  i.e. $\gamma_b=70.71$ . In panel (a) we present the function $\Gamma(k,\omega_B)$, while panel (b) shows contours of constant values of $\Gamma(k,\omega_B)$  in the $(k,\omega_B)$  plane.}
    \label{Fig5}
\end{figure}

Fig. \ref{Fig5} shows the dependence of the instability growth rate in the non-resonant (long-wavelength) limit, together with the growth rate at shorter wavelengths. In panel (a), we present the function  $\Gamma(k,\omega_B)$, while panel (b) shows contours of constant values  of $\Gamma(k,\omega_B)$  in the $(k,\omega_B)$  plane. We see that 
in the highly magnetized plasma with $\omega_B>>\omega_p$ the instability develops in the cyclotron resonance regime when the beam mode 
\begin{equation}
 \label{eq404}
(\omega-kc \beta_b)^2\approx\frac{\omega_B^{2}}{\gamma_b^2} 
\end{equation}
is in the resonance with the electromagnetic wave whose dispersion equation is given by Eq. (\ref{eq5}). This yields equation for resonance frequency 
\begin{equation}
 \label{405}
\omega_0\left(\beta_b \sqrt{1+\frac{\omega_p^2}{\omega_B^2-\omega_0^2}}\right)=\frac{\omega_B}{\gamma_b}. 
\end{equation}
Representing the unstable mode frequency as 
\begin{equation}
 \label{406}
\omega=\omega_0+\delta 
\end{equation}
with $\delta<<\omega_0$ and substituting it in Eq. (\ref{eq33}) we can find the instability growth rate $\Gamma=\Im[\omega]$ approximately equal to
\begin{equation}
 \label{407}
\Gamma\approx\frac{\omega_b}{2 \sqrt{\gamma_b}}.
\end{equation}

Using the relations obtained above, we can find the group velocity of the unstable resonant mode ($v_g=\partial_k\omega$). It is always less than the beam velocity. In the limit $\omega_B<< \omega_p$, it is approximately equal to $c\beta_b/2$, despite the fact that the beam is ultrarelativistic.

As can be seen, the electron–positron beam instability can generate a low-frequency electromagnetic mode. This mode can subsequently be amplified and frequency-upshifted upon reflection from the front of a relativistic shock wave propagating in a magnetized electron–positron pair plasma if the group velocity of the Alfvén-like mode is lower than the shock wave velocity, i.e., lower than the velocity of the refractive-index modulations, i. e. $c\beta_M$. Although, under these conditions, the beam–plasma mode propagates in the same direction as the shock wave in the laboratory frame, in the shock-rest frame it propagates toward the shock front and can therefore be reflected from it.

Fast electron–positron beams capable of exciting low-frequency, slow electromagnetic waves in the upstream region are naturally generated at collisionless shock fronts through a variety of acceleration mechanisms, including surfatron acceleration, gradient- and curvature-drift acceleration, and diffusive (first type Fermi acceleration process) acceleration in turbulent plasma. Comprehensive reviews of charged-particle acceleration are provided by 
Berezinskii et al. \cite{Berezinskii1990}, Hoshino et al. \cite{Hoshino1991}, Gallant et al.\cite{Galant1992}, Hoshino \cite{Hoshino2008}, Spitkovsky \cite{Spitkovsky2008}, Sironi and Spitkovsky \cite{Sironi2011}, Sironi et al. \cite{Sironi2015}, and the references cited therein.

From the standpoint of particle acceleration efficiency, perpendicular and oblique shocks are generally more favorable than parallel shocks in the strong-magnetization regime. In contrast, the photon-acceleration mechanism considered here is most effective for parallel shock propagation, which provides the largest frequency upshift of the electromagnetic wave. The consideration of photon acceleration in an oblique shock-wave configuration is beyond the scope of the present work and will be addressed in a future study.

\section{Photon acceleration as a theoretical model for Fast Radio Bursts}

Here, considering the photon accelerator as a mechanism for fast radio bursts, we relate the theoretical predictions obtained above to the typical parameters of FRB known from observations (see the review articles Popov et al. \cite{Popov2018}, Cordes and Chatterjee \cite{Cordes2019}, Petroff, Hessels, and Lorimer \cite{Petroff2022}, Zhang \cite{Zhang2023}). 

The FRB pulse duration is one millisecond  $\Delta t_{FRB}=10^{-3}$s, from which we can find the FRB pulse length,  $l_{FRB}=c\Delta t_{FRB}=3\times 10^7$cm. The carrier frequency of the FRB pulse ranges from $\nu_{FRB}=\omega_{FRB}/2 \pi=300\,$MHz to 8 GHz. The FRB energy spans from  $\mathcal{E}_{FRB}=10^{38}$erg to $10^{41}$erg.

In the photon accelerator, due to the double Doppler effect, the frequency of the precursor electromagnetic field is upshifted,  the longitudinal size compressed and the field amplitude increased by a factor approximately equal to $4\gamma_M^2$. 
If the FRB radiation is collimated within an angle approximately equal to $\theta=\gamma_M^{-1}$  the FRB energy should be lower by a factor of  $\approx\gamma_M^{-2}$.

Since the FRB frequency is about the Larmor frequency $\omega_{FRB}=\omega_B$  we can find the magnetic field in the region where the FRB is formed. It is equal to
\begin{equation}
 \label{eq41}
B_0=\frac{m_e\omega_B c}{e}=60\left(\frac{\omega_{FRB}}{10^9\text{Hz}} \right)  G.  
\end{equation}

Assuming that the beam instability saturates due to nonlinear effects when the perturbations reach an amplitude approximately equal to $B_0$, i.e., the precursor wave has approximately this amplitude, we can estimate the FRB amplitude as $\approx 4\gamma_M^2 B_0$.

Considering the FRB pulse length $l_{FRB}$ as the relativistically compressed version of the initial low-frequency pulse length $l_{FRB,0}$, we can estimate the latter as 
$l_{FRB,0} \approx 4\gamma_M^2 l_{FRB}$.

Using these relationships, we can write for the FRB energy.
\begin{equation}
 \label{eq42}
\mathcal{E}_{FRB}=4\gamma_M^2\frac{B_{FRB}^2}{4\pi}l_{FRB,0}^2 l_{FRB}=256\gamma_M^{10}\frac{B_0^2}{\pi}l_{FRB}^3,
\end{equation}
which yields the relativistic shock wave Lorentz factor 
\begin{equation}
 \label{eq43}
\gamma_M=16\left(\frac{\mathcal{E}_{FRB}}{10^{40}\text{erg}}\right)^{1/10} \left(\frac{\Delta t_{FRB}}{10^{-3}\text{s}}\right)^{-3/10}
\left(\frac{\omega_{FRB}}{10^{9}\text{Hz}}\right)^{-1/5}.
\end{equation}
Now we can find the size of the region where FRB is generated. It is 
$l_{FRB,0}=4\gamma_M^2l_{FRB}=3\times 10^{10}\text{cm}$, which is comparable with the size of the magnetar magnetosphere. If the magnetic field strength at the magnetar surface equals $B_M$, the magnetic field at the distance equal to  $l_{FRB,0}$ should be $B_0$. This requires consideration of  the twisted magnetosphere model (Beloborodov\cite{Beloborodov2017}, Voisin and Francez \cite{Voisin2025}) where the relationship between $B_M$ and $B_0$ is given by
\begin{equation}
\frac{B_0}{60\text{G}}=1.7\times 10^{12}\left(\frac{B_M}{10^{14}\text{G}}\right)\left(\frac{R_M}{10^6\text{cm}}\right)^{2+p}\left(\frac{l_{FRB}}{3\times 10^{10}\text{cm}}\right)^{-2-p},
\end{equation}
Where $R_M$  is the magnetar radius and $0\leq p \leq 1$  is the twist parameters. 
From this equation, we obtain the twist parameter equal to $p=0.73$  for $B_M=10^{14} \text{G}$ and  $p=0.95$ if   $B_M=10^{15} \text{G}$ .

Here, we consider an external relativistic shock generated either by ejecta produced during a starquake or by magnetic reconnection, with a characteristic isotropic-equivalent energy of $\mathcal{E}_{sh}\approx 10^{44}$ erg, as expected for a powerful magnetar flare. Such a shock can propagate into the upstream medium and interact with the low-frequency waves generated by the precursor. This interaction can intensify the waves and upshift their frequency, causing them to emerge as an FRB. We adopt a shock Lorentz factor of $\gamma_M=16$.

The precursor wavenumber can be obtained from Eq. (\ref{eq39}) and is given by
$k_p=\omega_B/(c\sqrt{\gamma_b})$.
The precursor wave is subsequently intensified and its frequency is upshifted to $4\gamma_M^2 k_p c$, which is equal to the cyclotron frequency $\omega_B$. This condition yields the Lorentz factor of the relativistic electron-positron beam,
$\gamma_b=4\gamma_M^2\approx 10^3$.

The energy of the beam,
$\mathcal{E}_{b}=l_0^3 m_e c^2 \gamma_b n_b$,
must be sufficiently large to supply the energy required for the precursor waves, $B_0^2l_0^3/4\pi$. We therefore obtain the required electron beam density,
$n_b\approx 4\times10^7\ {\rm cm}^{-3}$,
which is substantially lower than the typical electron density in the outer regions of the magnetar magnetosphere (see the discussion of magnetospheric parameters and physical processes in Qu, Kumar, \& Zhang \cite{Qu2022}).

The corresponding relativistic beam energy,
$\mathcal{E}_{b}=\mathcal{E}_{FRB}/4\gamma_M^2$,
is approximately three orders of magnitude smaller than the FRB energy. As the instability develops, the energy of the relativistic electrons is depleted over a characteristic distance determined by the instability growth rate,
$\Gamma=\omega_b/\sqrt{\gamma_B}\approx 3\times10^2\ {\rm s}^{-1}$,
which corresponds to a timescale substantially shorter than the typical duration of FRB formation.

The timescale for electron radiation losses in the outer magnetar magnetosphere (see, e.g., Qu, Kumar, \& Zhang \cite{Qu2022} and Beloborodov \cite{Beloborodov2022}) is of order
$\tau_{\rm rad}\approx 3/(2r_e\gamma_b\omega_B^2)
=(3/2)(10^3/\gamma_b)(10,{\rm GHz}/\omega_B)^2\approx1\ {\rm s}$, where $r_e=e^2/m_e c^2\approx 2.8\times 10^{-13}$ cm is the classical electron radius.
This timescale is longer than the characteristic timescale of the instability development, indicating that radiative losses do not significantly affect the instability on the relevant timescale.

\section{Conclusion}

We propose a physical mechanism for Fast Radio Bursts (FRBs) based on the concept of photon acceleration. As in several previously published studies, the central role is played by a relativistic shock wave. We consider a relativistic shock propagating along the magnetic field in an electron–positron pair plasma, a configuration that may be realized in a magnetar magnetosphere. Density modulations of the electron–positron plasma at the shock front generate corresponding perturbations of the refractive index that propagate at relativistic velocity. The interaction of these moving refractive-index perturbations with a relatively low-frequency electromagnetic field converts the latter into an intensified high-frequency electromagnetic wave. The low-frequency field is assumed to originate from electromagnetic precursors produced by the instability of an electron–positron beam accelerated by the shock wave and interacting with the upstream plasma.

\begin{figure}[!htbp]
    \centering
    \includegraphics[width=\linewidth]{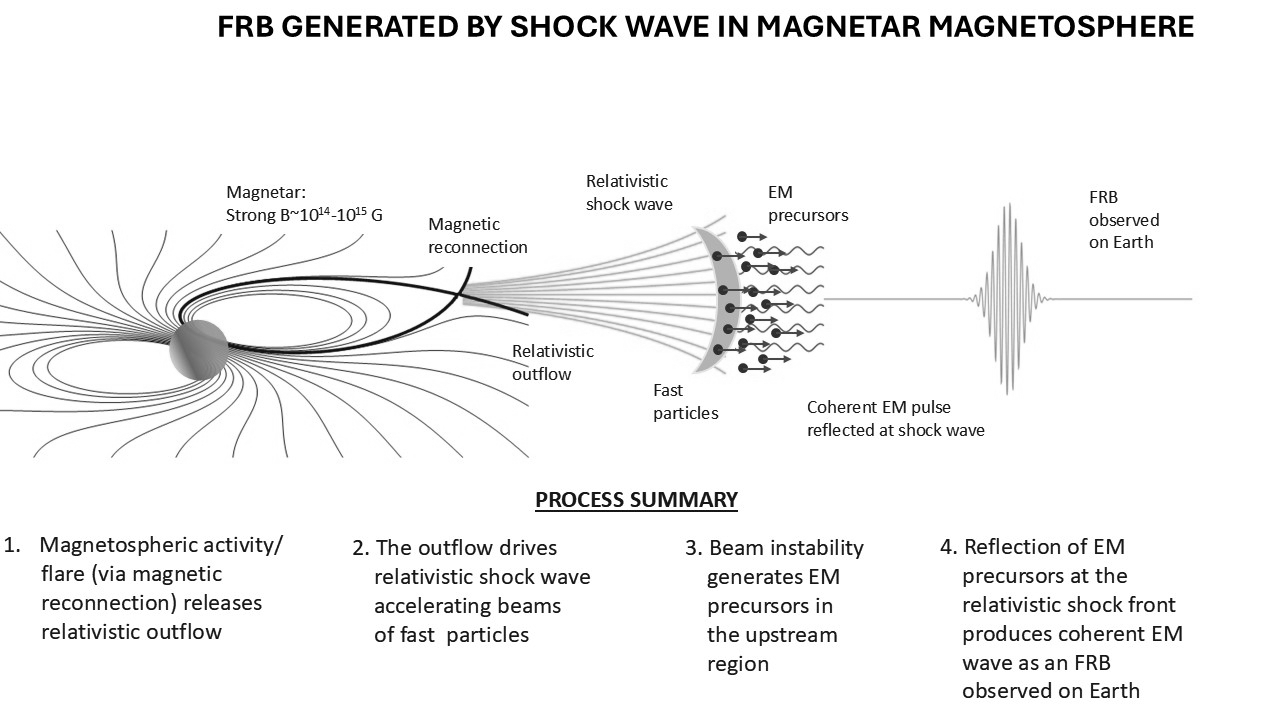}
    \caption{ 
    Summary of the process resulting in the FRB generation at the shock wave in magnetar magnetosphere. Magnetospheric activity in a magnetar triggers a flare that releases magnetic energy through magnetic reconnection, producing a relativistic plasma outflow. This outflow drives a relativistic shock wave into the surrounding ambient plasma. Charged particles are accelerated at the shock front and subsequently generate low-frequency electromagnetic precursors in the upstream region. Upon reflection from the relativistic shock, these precursors are converted into a coherent, high-intensity electromagnetic pulse. This pulse propagates through space and is observed on Earth as a fast radio burst. }
    \label{Fig6}
\end{figure}

The concept of photon acceleration, formulated by Wilks et al. \cite{Wilks}, can be traced back to the seminal work of Einstein \cite{Einstein}. In that study, A. Einstein considered the reflection of an electromagnetic wave from a mirror moving at relativistic velocity to illustrate the fundamental principles of special relativity. He showed that such a reflection results in an increase in both the frequency and amplitude of the electromagnetic wave.

The theory of photon acceleration employed in the present paper is formulated within the framework of geometric optics, which describes the limiting cases of complete reflection or transmission of the wave. An important extension of this concept was proposed by Bulanov, Esirkepov and Tajima \cite{Bulanov2003}, who considered relativistic plasma mirrors formed by nonlinear plasma waves. Wave breaking in such structures produces singular electron-density distributions that enable partial reflection of electromagnetic radiation. Further developments and applications of this concept are reviewed by Mourou, Tajima and Bulanov \cite{Mourou2006}; Bulanov et al. \cite{Bulanov2013} and by Kato, Mima and Bulanov \cite{Kato2026}.
The equations of geometric optics, written in Hamiltonian–Lagrangian form, possess a conserved Jacobi integral. 
Analysis of this invariant allows one to determine the properties of electromagnetic waves reflected from the shock front.

Extending the photon-acceleration framework to magnetized plasma, as required for astrophysical and space-plasma applications, leads to qualitatively new features. In particular, the reflectivity of a relativistically moving electron–positron plasma slab is significantly higher than in the unmagnetized case. Furthermore, the theory predicts that the frequency upshift of initially low-frequency waves is bound from above by the electron Larmor frequency.

Taking this limitation into account, together with the dependence of the frequency upshift and wave-amplitude amplification on the shock Lorentz factor, and using the observed energies, frequencies, and durations of FRBs, we infer the parameters required for the underlying physical mechanism. 
Remarkably, these parameters are consistent with those expected in magnetar magnetospheres, including the characteristic spatial scales, magnetic-field strengths, and relativistic shock-wave Lorentz factors. They are also compatible with contemporary models of twisted magnetar magnetospheres (Beloborodov\cite{Beloborodov2017}; Voisin and Francez \cite{Voisin2025}).

A summary of the process is illustrated in Fig. \ref{Fig6}.\\


\textbf{Acknowledgements}
The authors would like to thank  Masahiro Hoshino, Remo Ruffini, and Bing Zhang for fruitful discussions. AGRT was supported by National Science Foundation Grant No. 2512014.\\


\begin{thebibliography}{}

\bibitem[1]{Lorimer} Lorimer, D. R., Bailes, M., McLaughlin, et al. 2007, Science 318 (5851), 777

\bibitem[2]{LorimerRev} Lorimer, D. R., Bailes, M., McLaughlin, 2024, A \& A 369, 59

\bibitem[3]{Petroff2022} Petroff, E., Hessels, J. W. T., \& Lorimer, D. R. 2022, Astron. Astrophys. Rev. 30, 2

\bibitem[4]{Lyubarsky2014} Lyubarsky, Y. 2014, MNRAS, 442, L9 

\bibitem[5]{Beloborodov2017} Beloborodov, A. M. 2017, ApJL 843, L26

\bibitem[6]{Popov2018} Popov, S. B., Postnov, K. A., \& Pshirkov, M. S. 2018, Physics Uspekhi 61, 965

\bibitem[7]{Zhang2020} Zhang, B., 2020, Nature 587, 45

\bibitem[8]{Lyutikov2021} Lyutikov, M. 2021, ApJ 922, 166 

\bibitem[9]{Zhang2023} Zhang, B., 2023, Rev. Mod. Phys. 95, 035005

\bibitem[10]{Colgate1971} Colgate, S. A., \& Noerdlinger, P. D. 1971, ApJ, 165, 509

\bibitem[11]{Iwamoto2} Iwamoto, M., Matsumoto, Y., Amano, T., et al. 2024, Phys. Rev. Lett. 132, 035201

\bibitem[12]{Yalinewich} Yalinewich, A. \& Pen, U.-L. 2022, MNRAS 515, 5682

\bibitem[13]{Einstein} Einstein, A. 1905, Ann. Phys. (Leipzig) 17, 891

\bibitem[14]{Deng} Deng, D.-C., \& Wu, H.-C. 2026, MNRAS 546, 1 

\bibitem[15]{Wilks} Wilks, S. C., Dawson, J. M., Mori, et al. 1989, Phys. Rev. Lett. 62, 2600

\bibitem[16]{Mendonca2001} Mendonça, J. T. 2001, Theory of Photon Acceleration. (Bristol and Philadelphia: Institute of Physics Publishing)

\bibitem[17]{Eliezer2005} Eliezer, S., Mendonça, J. T., Bingham, R., \& Norreys, P. 2005, Physics Letters A 336, 390

\bibitem[18]{Bulanov2026} Bulanov, S. V., Bulanov, S. S., Esirkepov, T. Zh., et al. 2026, New J. Phys. 28, 064302

\bibitem[19]{Hadjisolomou2023} Hadjisolomou, P., Shaisultanov, R., Jeong, T. M. et al. 2023, Phys. Rev. Research 5, 043153

\bibitem[20]{Iwamoto1} Iwamoto, M., Amano, T., Hoshino, M., et al. 2019, ApJL 883, L35

\bibitem[21]{Marcowith} Marcowith, A.,et al. 2016, Reports on Progress in Physics, 79, 046901

\bibitem[22]{Plotnikov2019} Plotnikov, I. \& Sironi, L. 2019, MNRAS 485,3816

\bibitem[23]{Vanthieghem2025} Vanthieghem, A., \& Levinson, A., 2025, Phys. Rev. Lett. 134, 035201

\bibitem[24]{Mahlman2026} Mahlman, J. F., Eskildsen, L., Vanthieghem, A., et al. 2026, ApJL 999, L15

\bibitem[25]{Kazimura1998} Kazimura, Y., Sakai, J.-I., Neubert, T., \& Bulanov, S. V. 1998, ApJL 498, L183

\bibitem[26]{Medvedev1999} Medvedev, M. V., \& Loeb, A. 1999, ApJ, 526, 697

\bibitem[27]{Beloborodov2013} Beloborodov, A. M., 2013, ApJ, 777, 114 

\bibitem[28]{Golant25} Golant, R., Vanthieghem, A., Grošelj, D., \& Sironi, L. 2025, ApJ 986, 211

\bibitem[29]{Achatz1990} Achatz, U., Lesch, H., \& Schlickeiser, R. 1990, A\&A 233, 391

\bibitem[30]{Achatz1993} Achatz, U., \& Schlickeiser, R. 1993, A\&A 274, 165

\bibitem[31]{Lifshitz1981} Lifshitz, E. M. \& Pitaevskii, L. P., Physical Kinetics, Course of Theoretical Physics, Vol. 10 (Pergamon Press, Oxford, 1981)

\bibitem[32]{Berezinskii1990} Berezinskii, V. S., Bulanov, S. V., Ginzburg, V. L., Dogiel, V. A. \& Ptuskin, V. S. Astrophysics of cosmic rays. (North Holland Publ. Co. Elsevier Sci. Publ. Amsterdam, 1990).

\bibitem[33]{Hoshino1991} Hoshino, M., \& Arons, J. 1991, Phys. Fluids B 3, 818

\bibitem[34]{Galant1992} Gallant, Y. A., Hoshino, M., Langdon, A. B., Arons, J., \& Max, C. E. 1992, ApJ 391, 73

\bibitem[35]{Hoshino2008} Hoshino, M., 2008, ApJ 672, 940

\bibitem[36]{Spitkovsky2008} Spitkovsky, A. 2008, ApJL 682, L5 

\bibitem[37]{Sironi2011} Sironi, L., \& Spitkovsky, A. 2011, ApJ 741, 39

\bibitem[38]{Sironi2015} Sironi, L, Keshet, U., \& Lemoine, M. 2015, Space Sci. Rev. 191, 519

\bibitem[39]{Cordes2019} Cordes, J. M., \& Chatterjee, S. 2019, Annu. Rev. Astron. Astrophys. 57, 417

\bibitem[40]{Voisin2025} Voisin, G., \& Francez, T., 2025, A\&A 702, A55

\bibitem[41]{Qu2022} Qu, Y., Kumar, P., \& Zhang, B., 2022, MNRAS 515, 2020

\bibitem[42]{Beloborodov2022} Beloborodov, A. M., 2022, Phys. Rev. Lett. 128, 255003

\bibitem[43]{Bulanov2003} Bulanov, S. V., Esirkepov, T., \& Tajima, T. 2003, Phys. Rev. Lett. 91, 085001

\bibitem[44]{Mourou2006} Mourou, G., Tajima, T., \& Bulanov, S. V. 2006, Rev. Mod. Phys. 78, 309

\bibitem[45]{Bulanov2013}Bulanov, S. V., Esirkepov, T. Zh., Kando, M. et al. 2013, Physics Uspekhi 56, 429


\bibitem[46]{Kato2026} Kato, Y., Mima, K., \& Bulanov, S. V., High power laser and plasma science (Springer Series on Atomic, Optical, and Plasma Physics; SSAOPP, volume 130, 2026)


\end{thebibliography}
\end{document}